\documentclass[final,authoryear,1p,12pt]{elsarticle}

\usepackage{graphicx}
\usepackage{epstopdf}
\usepackage{booktabs}
\usepackage{natbib}
\usepackage{fancyhdr}
\usepackage{amsmath}
\usepackage{amsfonts}
\usepackage{multirow}
\usepackage{lscape}
\usepackage{setspace}
\usepackage{color}
\usepackage{hyperref}
\usepackage{color,soul}
\sethlcolor{green}
\usepackage{ifthen}
\usepackage[latin1]{inputenc}
\usepackage{titletoc}
\usepackage{eso-pic}
\definecolor{123}{rgb}{.9,.9,.9}
\hypersetup{
colorlinks=false,
citecolor=blue,
linkbordercolor={1 1 1}, 
citebordercolor={1 1 1},
urlbordercolor={1 1 1}
}
\usepackage{bbm}
\usepackage{amssymb}
 \usepackage{amsthm}
 \usepackage{xcolor}

\DeclareMathOperator{\plim}{plim}

\begin{document}
\begin{frontmatter}

\title{Modeling and forecasting exchange rate volatility in time-frequency domain\tnoteref{label1} }

\author[ies,utia]{Jozef Barunik\corref{cor2}}
\author[ies,utia]{Tomas Krehlik}
\author[ies,utia]{Lukas Vacha}

\address[ies]{Institute of Economic Studies, Charles University, \\ Opletalova 21, 110 00, Prague,  Czech Republic}
\address[utia]{Institute of Information Theory and Automation, Academy of Sciences of the Czech Republic, Pod Vodarenskou Vezi 4, 182 00, Prague, Czech Republic}
\cortext[cor2]{Corresponding author, Tel. +420(776)259273, Email address: barunik@utia.cas.cz}
\tnotetext[label1]{We are grateful to the editor Lorenzo Peccati and two anonymous referees for many useful comments and suggestions, which greatly improved the paper. We are also grateful to David Veredas and Karel Najzar, and seminar participants at the Modeling High Frequency Data in Finance 3 in New York (July 2011) and Computational and Financial Econometrics in Oviedo (December 2012) for many useful discussions.  The research leading to these results has received funding from the European Union's Seventh Framework Programme (FP7/2007-2013) under grant agreement No. FP7-SSH- 612955 (FinMaP). Support from the Czech Science Foundation under the 13-32263S project is gratefully acknowledged.}

\begin{abstract}
{\small This paper proposes an enhanced approach to modeling and forecasting volatility using high frequency data. Using a forecasting model based on Realized GARCH with multiple time-frequency decomposed realized volatility measures, we study the influence of different timescales on volatility forecasts. The decomposition of volatility into several timescales approximates the behaviour of traders at corresponding investment horizons. The proposed methodology is moreover able to account for impact of jumps due to a recently proposed jump wavelet two scale realized volatility estimator. We propose a realized Jump-GARCH models estimated in two versions using maximum likelihood as well as observation-driven estimation framework of generalized autoregressive score. We  compare forecasts using several popular realized volatility measures on foreign exchange rate futures data covering the recent financial crisis. Our results indicate that disentangling jump variation from the integrated variation is important for forecasting performance. An interesting insight into the volatility process is also provided by its multiscale decomposition. We find that most of the information for future volatility comes from high frequency part of the spectra representing very short investment horizons. Our newly proposed models outperform statistically the popular as well conventional models in both one-day and multi-period-ahead forecasting.}
\end{abstract}
\end{frontmatter}

\section{Introduction}

In contrast to the conventional framework of a generalized autoregressive conditional heteroscedasticity (GARCH) model, volatility is directly observed and can be used for forecasting when high frequency data are applied.\footnote{A vast quantity of literature on several aspects of estimating volatility using high frequency data is nicely surveyed by \cite{macaleer}.} While \cite{hansen2005forecast} argue that GARCH(1,1) can hardly be beaten by any other model, recent active research shows that with help of high frequency measures, we can improve the volatility  forecasts significantly. \cite{mcmillan2012daily} for example utilize intra-day data and show that we can obtain forecasts superior to forecasts from GARCH(1,1). \cite{louzis2013role} assesses the informational content of alternative realized volatility estimators using Realized GARCH in Value-at-Risk prediction. We aim to extend this line of research by investigating the importance of disentangling jump variation and integrated variance in recently developed framework, which combines appeal of a widely used GARCH(1,1) and high frequency data. Moreover, we employ recently developed multiscale estimators which decompose volatility into several investment horizons and allow us to study the influence of intraday investment horizons on the volatility forecasts.

Traders on financial markets make their decisions over different horizons, for example, minutes, hours, days, or even longer such as months and years \citep{lebaron2001stochastic,Ramsey2002,genccay2005multiscale,corsi2009}. Nevertheless, majority of the empirical literature studies the relationships in the time domain only aggregating the behavior across all investment horizons. A notable exception is the Heterogenous Autoregressive approach (HAR) for realized volatility proposed by \cite{corsi2009}. Although staying in the time domain solely, \cite{corsi2009} builds his model on the idea of the investors' heterogeneity. 

In our work, we ask if wavelet decomposition can provide better insight into the foreign exchange volatility modeling and forecasting. Wavelets are often successfully used as a de-noising tool \citep{haven2012noising,sun2012new}. One particularly appealing feature of wavelets is that they can be embedded into stochastic processes, as shown by \cite{antoniou1999}. Thus we can conveniently use them to extend the theory of realized measures to obtain decomposed volatility as shown by \cite{fanwang2008}, or \cite{barunik}. One of the common issues with the interpretation of wavelets in economic applications is that they are filter, thus they can hardly be used for forecasting in econometrics. Models based on wavelets are often outperformed by simple benchmark models, as shown by \cite{fernandez2008traditional}. Rather, they can provide a useful ``lens" into the spectral properties of the time series. Our wavelet-based estimator of realized volatility uses wavelets only to decompose the daily variation of the returns using intraday information, hence the problem with forecasts is no longer an issue. As wavelets are used to measure realized volatility at different investment horizons, we can construct a forecasting model based on the wavelet decomposed volatility conveniently.

Several attempts to use wavelets in the estimation of realized variation have emerged in the past few years. \cite{hoglunde2003} were the first to suggest a wavelet estimator of realized variance. \cite{capobianco2004}, for example, proposes to use a wavelet transform as a comparable estimator of quadratic variation. \cite{subbotin2008} uses wavelets to decompose volatility into a multi-horizon scale. One exception is the work of \cite{fanwang2008}, who were the first to use the wavelet-based realized variance estimator together with the methodology for estimation of jumps. In \cite{barunik}, we revisit and extend this work and using large Monte Carlo study we show that this estimator improves forecasting of the volatility substantially when compared to other estimators. Moreover, in \cite{barunik} we attempt to use the estimators to decompose stock market volatility into several investment horizons in a non-parametric way.

Motivated by previous results, this paper focuses on proposing a model which will improve the modeling and forecasting of foreign exchange volatility. Similarly to \cite{lanne2007}, \cite{ABH2011}, and \cite{sevi2014forecasting} we use the decomposition of the quadratic variation with the intention of building a more accurate forecasting model. Our approach is very different though, as we use wavelets to decompose the integrated volatility into several investment horizons and jumps. Moreover, we employ recently proposed realized GARCH framework of \cite{Hansen2011}. In contrast to popular HAR framework of \cite{corsi2009}, Realized GARCH allows to model jointly returns and realized measures of volatility, while key feature is a measurement equation that relates the realized measure to the conditional variance of returns. In addition, we benchmark our approach to several measures of realized volatility and jumps, namely realized volatility estimator proposed by \cite{abdl2003}, the bipower variation estimator of \cite{barndorff2004}, the median realized volatility of \cite{andersen2012jump}, and finally jump wavelet two-scale realized variance (JWTSRV) estimator of \cite{barunik} in the framework of Realized GARCH, and we find significant differences in volatility forecasts, while our JWTSRV estimator brings the largest improvement. We use Realized GARCH models of \cite{Hansen2011} as well as realized GAS of \cite{huang2014generalized} based on the observation-driven estimation framework of generalized autoregressive score models to build a realized Jump-GARCH modeling strategy. In addition, we also utilize Realized GARCH with multiple realized measures \citep{hansen2012exponential} to build a time-frequency model for forecasting volatility.

The main contribution of the paper is thus threefold. First, we propose several model extensions to utilize jumps in the popular Realized GARCH frameworks, as well as build time-frequency model for forecasting volatility. Second, we use several  popular measures as a benchmark to our time-frequency model. Third, we bring interesting empirical comparison of all frameworks in multiple-period-ahead forecasting exercise. We show that the most important information influencing the future volatility in foreign exchange is carried by the high frequency part of the spectra representing very short investment horizons of 10 minutes. This decomposition gives us an interesting insight into the volatility process. Our newly proposed time-frequency models and Jump-GARCH models outperforms the existing modeling strategies significantly.

\section{Theoretical framework for time-frequency decomposition of realized volatility}

While most time series models are naturally set in the time domain, wavelet transform help us to enrich the analysis of quadratic variation by the frequency domain. Traders of the foreign exchange markets are operating with heterogeneous expectations, ranging from minutes to days, or even weeks and months. Hence volatility dynamics should be understood not only in time but at different investment horizons as well. In this section, we introduce a multiscale estimator that will allow these features and is moreover able to separate the continuous part of the price process containing noise from the jump variation. We will briefly introduce general ideas of constructing the estimator, while for the details necessary to understand the derivation of the estimator using wavelet theory, we refer to \cite{barunik}. In addition, we introduce several other estimators commonly used in the literature, which will serve as a benchmarks to us in the empirical application.

In the analysis, we assume that the latent logarithmic asset price follows a standard jump-diffusion process contamined with microstructure noise. Let  $y_t$  be the observed logarithmic prices evolving over $0 \le t \le T$, which will have two components; the latent, so-called ``true log-price process",  $dp_t=\mu_t dt+\sigma_t dW_t+\xi_t dq_t$, and zero mean $i.i.d.$ microstructure noise, $\epsilon_t$, with variance $\eta^2$. In a latent process, $q_t$ is a Poisson process uncorrelated with $W_t$, and the magnitude of the jump, denoted as $J_l$, is controlled by factor $\xi_t \sim N(\bar\xi,\sigma^2_{\xi})$. Thus, the observed  price process is $y_t=p_t+\epsilon_t$.

The quadratic return variation over the interval $[t-h,t]$, for $0\le h \le t\le T$ associated with the price process $y_t$ can be naturally decomposed into two parts: integrated variance of the latent price process,  $IV_{t,h}$ and jump variation $JV_{t,h}$
\begin{equation}
\label{waveqvjump}
QV_{t,h}=\underbrace{\int_{t-h}^t \sigma_s^2 ds}_{\mbox{$IV_{t,h}$}}+ \underbrace{\sum_{t-h \le l \le t} J_l^2}_{\mbox{$JV_{t,h}$}} 
\end{equation}
As detailed by \cite{abdl2001} and \cite{barndorff2002a}, quadratic variation is a natural measure of variability in the logarithmic price process. A simple consistent estimator of the overall quadratic variation under the assumption of zero noise contamination in the price process is provided by the well-known realized variance, introduced by \cite{ab98}. The realized variance over $\left[t-h,t\right]$ can be estimated as  
\begin{equation}
\label{rv}
\widehat{QV}^{(RV)}_{t,h}=\sum_{k=1}^N \left(\Delta_k y_{t}\right)^2\hspace{1mm},
\end{equation}
where $\Delta_k y_{t}=y_{t-h+\left(\frac{k}{N}\right)h}-y_{t-h+\left(\frac{k-1}{N}\right)h}$ is the $k$-th intraday return in the $\left[t-h,t\right]$ and $N$ is the number of intraday observations. The estimator in Eq.(\ref{rv}) converges in probability to $IV_{t,h}+JV_{t,h}$ as $N\rightarrow\infty$ \citep{ab98, abdl2001,abdl2003, barndorff2001,barndorff2002a,barndorff2002}. 

While the observed price process $y_t$ is contamined with noise and jumps in real data, we need to account for this, as the main object of interest is the $IV_t$ part of quadratic variation. \cite{zhang2005} propose solution to the noise contamination by introducing the so-called two-scale realized volatility (TSRV henceforth) estimator. They adopt a methodology for estimation of the quadratic variation utilizing all of the available data using an idea of precise bias estimation. The two-scale realized variation over $\left[t-h,t\right]$ is measured by
\begin{equation}
\label{tsrv}
\widehat{QV}_{t,h}^{(TSRV)}=\widehat{QV}_{t,h}^{(average)}-\frac{\bar{N}}{N} \widehat{QV}_{t,h}^{(all)},
\end{equation}
where $\widehat{QV}_{t,h}^{(all)}$ is computed as in Eq. (\ref{rv}) on all available data and $\widehat{QV}_{t,h}^{(average)}$ is constructed by averaging the estimators $\widehat{QV}_{t,h}^{(g)}$ obtained on $G$ grids of average size $\bar{N}=N/G$ as $\widehat{QV}_{t,h}^{(average)}=\frac{1}{G}\sum_{g=1}^G \widehat{QV}_{t,h}^{(g)}$, 
where the original grid of observation times, $M=\{t_1,\dots,t_N\}$ is subsampled to $M^{(g)}$, $g=1,\dots,G$, where $N/G\rightarrow \infty$ as $N\rightarrow \infty$. For example, $M^{(1)}$ will start at the first observation and take an observation every 5 minutes, $M^{(2)}$ will start at the second observation and take an observation every 5 minutes, ets. Finally, we average these estimators through the subsamples, so the variation of the estimator is averaged.

The estimator in Eq. (\ref{tsrv}) provides the first consistent and asymptotic estimator of the quadratic variation of $p_t$ with rate of convergence $N^{-1/6}$. \cite{zhang2005} also provide the theory for optimal choice of $G$ grids, $G^*=c N^{2/3}$, where the constant $c$ can be set to minimize the total asymptotic variance. 

Since we are interested in decomposing quadratic variation into the integrated variance and jump variation component, we introduce a methodology for jump detection. Recent evidence from the volatility forecasting literature indicates that two sources of variation in the price process substantially differ and impact future volatility in different ways. Before introducing our estimator, we introduce two commonly used estimators of volatility and integrated variation, which will be used as benchmark in the empirical exercise. 

\cite{barndorff2004,barndorff2006} develop bipower variation estimator (BV), which can detect the presence of jumps in high-frequency data. The main idea of the BV estimator is to compare two measures of the integrated variance, one containing the jump variation and the other being robust to jumps and hence containing only the integrated variation part. In our work, we use the \cite{ABH2011} adjustment of the original \cite{barndorff2004} estimator, which helps render it robust to certain types of microstructure noise. The bipower variation over $[t-h,t]$ is defined by
 \begin{equation}
 \label{bv}
   \widehat{IV}^{(BV)}_{t,h}=\mu_1^{-2} \frac{N}{N-2} \sum_{k=3}^{N} |\Delta_{k-2} y_t|.|\Delta_k y_t|,
   \end{equation}
   where $\mu_a=\pi/2=E(|Z|^a),$ and $Z\sim N(0,1)$, $a \ge0$ and $ \widehat{IV}^{(BV)}_{t,h} \to \int_{t-h}^t \sigma_s^2 ds$.
Therefore, $ \widehat{IV}^{(BV)}_{t,h}$ provides a consistent estimator of the integrated variance. Although $ \widehat{QV}^{(RV)}_{t,h}$ provides a consistent estimator of the integrated variance plus the jump variation, the jump variation may be estimated consistently as the difference between the realized variance and the realized bipower variation
\begin{equation}
\plim_{N \to \infty} \left(\widehat{QV}^{(RV)}_{t,h}-\widehat{IV}^{(BV)}_{t,h}\right)=JV_{t,h}.
\end{equation}
Under the assumption of no jump and some other regularity conditions, \cite{barndorff2006} provide the joint asymptotic distribution of the jump variation.\footnote{Under the null hypothesis of no within-day jumps, $$ \mathcal{Z}^{(BV)}_{t,h}=\frac{\frac{\widehat{QV}^{(RV)}_{t,h}-\widehat{IV}^{(BV)}_{t,h}}{\widehat{QV}^{(RV)}_{t,h}}}{\sqrt{\left( \left(\frac{\pi}{2}\right)^2+\pi-5\right)\frac{1}{N} \max \left( 1,\frac{\widehat{TQ}^{(BV)}_{t,h}}{\left(\widehat{IV}^{(BV)}_{t,h} \right)^2} \right)}},$$ where 
$\widehat{TQ}^{(BV)}_{t,h}=N \mu_{4/3}^{-3}\left(\frac{N}{N-4}\right) \sum_{k=5}^{N} |\Delta_{k-4}y_t|^{4/3}|\Delta_{k-3}y_t|^{4/3} |\Delta_{k-2}y_t|^{4/3}$ is asymptotically standard normally distributed.} Using this theory, the contribution of the jump variation to the quadratic variation of the price process is measured by
\begin{equation}
\label{bvjump1}
\widehat{JV}^{(BV)}_{t,h}=\mathbbm{1}_{\{\mathcal{Z}^{(BV)}_{t,h}>\Phi_{\alpha}\}} \left (\widehat{QV}^{(RV)}_{t,h}-\widehat{IV}^{(BV)}_{t,h} \right),
\end{equation}
where $\mathbbm{1}_{\{\mathcal{Z}^{(BV)}_{t,h}>\Phi_{\alpha}\}}$ denotes the indicator function and $\Phi_{\alpha}$ refers to the chosen critical value from the standard normal distribution. The measure of integrated variance is defined as
\begin{equation}
\label{cbv}
\widehat{IV}^{(CBV)}_{t,h}=\mathbbm{1}_{\{\mathcal{Z}^{(BV)}_{t,h} \le \Phi_{\alpha}\}} \widehat{QV}^{(RV)}_{t,h}+\mathbbm{1}_{\{\mathcal{Z}^{(BV)}_{t,h}>\Phi_{\alpha}\}}\widehat{IV}^{(BV)}_{t,h},
\end{equation}
ensuring that the jump measure and the continuous part add up to the estimated variance without jumps.
Another popular estimator, which estimates the integrated volatility in the presence of jumps is the median realized volatility (MedRV), introduced by \cite{andersen2012jump}:
 \begin{equation}
 \label{medrv}
   \widehat{IV}^{(MedRV)}_{t,h}=\frac{\pi}{6-4 \sqrt{3}+\pi} \left(\frac{N}{N-2}\right) \sum_{k=3}^{N} \text{med} \left( |\Delta_{k-2}y_t|,|\Delta_{k-1}y_t|,|\Delta_{k}y_t| \right)^2.
   \end{equation}
Under the assumption of no jump and some other regularity conditions, \cite{andersen2012jump} provide the joint asymptotic distribution of the jump variation\footnote{Under the null hypothesis of no within-day jumps, $$ \mathcal{Z}^{(MedRV)}_{t,h}=\frac{\frac{\widehat{QV}^{(RV)}_{t,h}-\widehat{IV}^{(MedRV)}_{t,h}}{\widehat{QV}^{(RV)}_{t,h}}}{\sqrt{0.96\frac{1}{N} \max \left( 1,\frac{\widehat{TQ}^{(MedRV)}_{t,h}}{\left(\widehat{IV}^{(MedRV)}_{t,h} \right)^2} \right)}},$$ where 
$\widehat{TQ}^{(MedRV)}_{t,h}=\frac{3\pi N}{9\pi+72-52\sqrt{3}} \left(\frac{N}{N-2}\right) \sum_{k=3}^{N} \text{med} \left( |\Delta_{k-2}y_t|,|\Delta_{k-1}y_t|,|\Delta_{k}y_t| \right)^4$ is asymptotically standard normally distributed.} analogously to the BV estimator. The integrated variance and jump variation can be consistently estimated as
\begin{eqnarray}
\label{cmedrv}
\widehat{IV}^{(CMedRV)}_{t,h}&=&\mathbbm{1}_{\{\mathcal{Z}^{(MedRV)}_{t,h} \le \Phi_{\alpha}\}} \widehat{QV}^{(RV)}_{t,h}+\mathbbm{1}_{\{\mathcal{Z}^{(MedRV)}_{t,h}>\Phi_{\alpha}\}}\widehat{IV}^{(MedRV)}_{t,h}, \\
\widehat{JV}^{(MedRV)}_{t,h}&=&\mathbbm{1}_{\{\mathcal{Z}^{(MedRV)}_{t,h}>\Phi_{\alpha}\}} \left (\widehat{QV}^{(RV)}_{t,h}-\widehat{IV}^{(MedRV)}_{t,h} \right).
\end{eqnarray}

\subsection{Estimation of jumps and time-frequency realized variance}

\cite{fanwang2008} use a different approach to realized volatility measurement. They use wavelets in order to separate jump variation from the price process, as well as for estimation of the integrated variance on the jump--adjusted data. In addition, wavelet methodology offers decomposition of the estimated volatility into scales representing investment horizons. Therefore, we can observe how particular investment horizon contributes to the total variance. For a more detailed description of the wavelet transform see \ref{wt}. In the empirical section, we aim to study information content of investment horizons for volatility forecasting, thus we describe the wavelet jump detection and then introduce the wavelet estimator of integrated variance of  \cite{barunik}, which allows to decompose the volatility into several investment horizons.

As in the previous Section, we assume the sample path of the price process $y_t=p_t+\epsilon_t$ will have a finite number of jumps. Following results of \cite{wang95} on the wavelet jump detection and further extension of \cite{fanwang2008} to stochastic processes, we apply wavelet transform to detect jumps. Using effective localization properties of the wavelets, \cite{fanwang2008} show a way how to distinguish between continuous and jump part of the stochastic price process with $i.i.d.$ additive noise. They use the first scale of the discrete wavelet transform (the highest frequency) where the price process $p_t$ dominates the noise $\epsilon_t$ only close to a jump location, otherwise it is very small. In order to detect dominating part of the process $y_t$, \cite{fanwang2008} use the universal threshold of \cite{donoho}, defined as $D_t=d_t\sqrt{2 \log N}$, where $d_t=median\{|\mathcal{W}_{1,k}|\}/0.6745$ for $k\in[1,N]$ is a robust estimate of standard deviation. When the absolute value of the wavelet coefficient at the first scale is greater than a threshold $D_t$ then the noise $\epsilon_t$ is relatively small and the dominance of $p_t$ is caused by the jump part, therefore a jump is detected.

In the empirical part we adapt \cite{fanwang2008} procedure to the ``maximal overlap discrete wavelet transform" (MODWT). As a result a robust estimate of standard deviation has to be modified as: $d_t=\sqrt{2}\ median\{|\mathcal{W}_{1,k}|\}/0.6745$ for $k\in[1,N]$.\footnote{For more information about universal thresholds applied on the MODWT see \citep{PercivalWalden2000} and \cite{Gencay2002}.} Since we use the MODWT, we have $k$ wavelet coefficients at the first scale, which corresponds to number of intraday observations, i.e., $k=1,\dots,N$. In case the absolute value of the wavelet coefficient $|\mathcal{W}_{1,k}|$ is greater\footnote{Using the MODWT filters, we need to slightly correct the position of the wavelet coefficients to get the precise jump position, see \cite{PercivalMofjeld1997}.} then the threshold $D_t$ than a jump with size $\Delta_{k} J_{t}$ is detected as
\begin{equation}
\label{jdet}
\Delta_{k} J_{t} =\left(y_{t-h+\left(\frac{k}{N}\right)h}-y_{t-h+\left(\frac{k-1}{N}\right)h}
\right) \mathbbm{1}_{\left\{ |\mathcal{W}_{1,k}| >D_t\right\}}  \hspace{5mm} k\in [1,N].
\end{equation}

Following \cite{fanwang2008}, the jump variation over $[t-h,t]$ in the discrete time is estimated as the sum of squares of all the estimated jump sizes,
\begin{equation}
\label{jump}
\widehat{JV}_{t,h}=\sum_{k=1}^N \left(\Delta_{k} J_{t}\right)^2.
\end{equation} 
\cite{fanwang2008} prove that using (\ref{jump}), we are able to estimate the jump variation from the process consistently with the convergence rate of $N^{-1/4}$. 

Having precisely detected jumps, we proceed to jump adjustment of the observed price process $y_t$ over $\left[t-h,t\right]$. We adjust the data for jumps by subtracting the intraday jumps from the price process as:
\begin{equation}
\Delta_{k}y_t^{(J)}=\Delta_ky_t-\Delta_{k} J_{t}, \hspace{5mm} k=1,\dots N,
\end{equation}
where $N$ is the number of intraday observations.

Finally, the volatility can be computed using the jump-adjusted wavelet two-scale realized variance (JWTSRV) estimator on the jump adjusted data $\Delta_{k}y_t^{(J)}$.  The estimator utilizes the TSRV approach of \cite{zhang2005} as well as the wavelet jump detection method. Another advantage of the estimator is, that it decomposes the integrated variance into $J^m+1$ components, therefore we are able to study the dynamics of volatility at various investment horizons. Following \cite{barunik}, we define the JWTSRV estimator over $\left[t-h,t\right]$, on the jump-adjusted data as: 
\begin{equation}
\label{jwtsrv}
\widehat{IV}_{t,h}^{(JWTSRV)}=\sum_{j=1}^{J^m+1}\widehat{IV}_{j,t,h}^{(JWTSRV)}=\sum_{j=1}^{J^m+1}\left(\widehat{IV}_{j,t,h}^{(average)}-\frac{\bar{N}}{N} \widehat{IV}_{j,t,h}^{(all)}\right),
\end{equation}
where $\widehat{IV}_{j,t,h}^{(average)}=\frac{1}{G} \sum_{g=1}^G  \sum_{k=1}^N \left(\mathcal{W}_{j,k}^{(g)}\right)^2 $ is obtained from wavelet coefficient estimates on a grid of size $\bar{N}=N/G$, and $\widehat{IV}_{j,t,h}^{(all)}=\sum_{k=1}^N \left(\mathcal{W}_{j,k}\right)^2$ is the wavelet realized variance estimator at a scale $j$ on all the jump-adjusted observed data, $\Delta_{k}y_t^{(J)}$. $\mathcal{W}_{j,k}$ denotes the MODWT wavelet coefficient at scale $j$ with position $k$ obtained over $\left[t-h,t\right]$.

\cite{barunik} show that the JWTSRV is consistent estimator of the integrated variance as it converges in probability to the integrated variance of the process $p_t$, and they test the small sample performance of the estimator in a large Monte Carlo study. The JWTSRV is found to be able to recover true integrated variance from the noisy process with jumps very precisely. Moreover, the JWTSRV estimator is also tested in forecasting exercise, which confirms to improve forecasting of the integrated variance substantially.

\subsubsection{Bootstrapping the jump test using JWTSRV}
Although \cite{fanwang2008} showed the effectiveness of the wavelet jump detection, distribution properties of the estimated jump variation, and hence any test statistic stay unknown. In order to test for the presence of jumps using JWTSRV estimator, we propose to use the bootstrap test. Main reason for bootstrapping the jump test is that consistent estimator for the integrated quarticity is not analytically available for JWTSRV estimator. More importantly, finite sample properties of the jump tests based on functions of realized volatility estimators can be considerably improved using bootstrap, as noted by \cite{dovonon2014bootstrapping}. 

In order to obtain the bootstrapped distribution of test statistic under the assumption of no jumps, we generate $k$ intraday returns using estimated integrated part of the quadratic variation as $\Delta_k y^*_{t}=\sqrt{(1/k)\widehat{IV}_{t}^{(JWTSRV)}}\eta_{i,t}$, with $\eta_{i,t} \sim N(0,1)$ generated independently. $\widehat{QV}^{(RV^*)}_{t}$ and $\widehat{QV}^{(JWTSRV^*)}_{t}$ are then estimated on the given day $t$. Generating $b=1,\ldots,B$ realizations, we obtain
\begin{equation}
 \mathcal{Z}^{*}_{t,h}(b)=\frac{\widehat{QV}^{(RV^*)}_{t,h}-\widehat{IV}^{(JWTSRV^*)}_{t,h}}{\widehat{QV}^{(RV^*)}_{t,h}},
\end{equation}
which can be used to construct a bootstrap statistic to test the null hypothesis of no jumps as:
\begin{equation}
 \mathcal{Z}^{(JWTSRV)}_{t,h}=\frac{\frac{\widehat{QV}^{(RV^*)}_{t,h}-\widehat{IV}^{(JWTSRV^*)}_{t,h}}{\widehat{QV}^{(RV^*)}_{t,h}}-E(\mathcal{Z}^{*}_{t,h}(b))}{\sqrt{Var(\mathcal{Z}^{*}_{t,h}(b))}}.
\end{equation}
The bootstrap expectation and variance both depend on the data. We will rely on the assumptions of \cite{dovonon2014bootstrapping}, who show that under general conditions, this statistics will be normally distributed with limiting variance one, although they provide this result for the BV estimator. While we leave the rigorous treatment of this approach with JWTSRV estimator for the future work, we have studied the properties of the bootstrap test using simulations, which are available upon request from authors. 

The integrated variance and jump variation can then be consistently estimated as
\begin{eqnarray}
\label{cjwtsrv}
\widehat{IV}^{(CJWTSRV)}_{t,h}&=&\mathbbm{1}_{\{\mathcal{Z}^{(JWTSRV)}_{t,h} \le \Phi_{\alpha}\}} \widehat{QV}^{(RV)}_{t,h}+\mathbbm{1}_{\{\mathcal{Z}^{(JWTSRV)}_{t,h}>\Phi_{\alpha}\}}\widehat{IV}^{(JWTSRV)}_{t,h},\\
\widehat{JV}^{(JWTSRV)}_{t,h}&=&\mathbbm{1}_{\{\mathcal{Z}^{(JWTSRV)}_{t,h}>\Phi_{\alpha}\}} \left (\widehat{QV}^{(RV)}_{t,h}-\widehat{IV}^{(JWTSRV)}_{t,h} \right).
\end{eqnarray}

\section{A forecasting model based on decomposed integrated volatilities and jumps \label{forecastuni}}

Similarly to \cite{lanne2007}, \cite{ABH2011}, and \cite{sevi2014forecasting}, we use the decomposition of the quadratic variation with the intention of building a more accurate forecasting model. Our approach is very different though, as we use wavelets to decompose the integrated volatility into several investment horizons and jumps first. Then, we employ recently proposed Realized GARCH framework of \cite{Hansen2011} and its variants. Realized GARCH allows to model jointly returns and realized measures of volatility, while key feature is a measurement equation that relates the realized measure to the conditional variance of returns. We use the decomposed realized measures in the Realized GARCH, and expect that our modification will result in better in-sample fits of the data as well as out-of-sample forecasts. For comparison, we also use other estimators and study how they improve the forecasting ability of Realized GARCH.

\subsection{Realized Jump GARCH framework for forecasting}
The key object of interest in GARCH family is the conditional variance, $h_t=var(r_t|\mathcal{F}_{t-1})$, where $r_t$ is a time series of returns. While in a standard GARCH(1,1) model the conditional variance, $h_t$ is dependent on its past $h_{t-1}$ and $r^2_{t-1}$, \cite{Hansen2011} propose to utilize realized measures of volatility and make $h_t$ dependent on them as well. The authors introduce so-called measurement equation which ties the realized measure to latent volatility. The general framework of Realized GARCH$(p,q)$ models is well connected to existing literature in \cite{Hansen2011}. Here, we restrict ourselves to the simple log-linear specification of Realized GARCH$(1,1)$ with Gaussian innovations which we will use to build our model. 

Realized GARCH makes use of realized measures of volatility to help forecast the latent volatility process. In the previous sections, we have motivated several estimators, which allow us to disentangle continuous part and jump part of the quadratic return variation. While both parts may carry important information about future volatility, we propose a modified framework, which includes both. 

There are essentially two possible treatments of jumps in the Realized GARCH framework, depending on the belief about its endogenous or exogenous nature. In a large study, \cite{chatrath2014currency,lahaye2011jumps} show that currency jumps can be explained by U.S. macro announcements using the realized measures. This provides a good empirical evidence about the exogenous nature of jump arrivals. By addition of estimated jumps into the variance equation, we propose a Realized Jump-GARCH(1,1) model (Realized J-GARCH) given by
\begin{eqnarray}
\label{RJ-G}
r_t &=&\sqrt{h_t} z_t, \\
\log(h_t)&=& \omega+\beta \log(h_{t-1})+\gamma \log(x_{t-1})+\gamma_J \log(1+JV_{t-1}),\\
\log(x_t)&=& \xi+\phi \log(h_t)+\tau_1 z_t+\tau_2 z^2_t+u_t,
\end{eqnarray}
where $r_t$ is the return, $x_t$ and $JV_t$ are estimated continuous and jump components of quadratic variation using BV, MedRV, or JWTSRV realized measures, and $z_t$ and $u_t$ come from Gaussian normal distribution and are mutually independent. $\tau_1z_t+\tau_2 z_t^2$ is leverage function. If jumps have a significant impact on volatility forecasts, $\gamma_J$ coefficient should be significantly different from zero. For $\gamma_J=0$, the model reduces to the original Realized GARCH.

\cite{Hansen2011} motivate possibility of obtaining feasible multi-period-ahead forecasts as one of the main advantages of this framework. Multi-period-ahead predictions with the Realized GARCH model are straightforward with the use of vector autoregression structure for $log(h_t)$ and $log(x_t)$. In this paper, we follow this simple approach. In order to obtain multiple-period-ahead forecasts, we need to include jump component to the forecasting structure. Once we are treating jumps as an exogenous process, we simply use the ARMA structure for the $\log(1+JV_{t-1})$, which allows to obtain the multiple-period-ahead forecasts analogously to the Realized GARCH model. 

\subsection{Realized GARCH model based on decomposed integrated volatiles}

In addition to jumps, we also utilize decomposition of JWTSRV to see which investment horizon has impact on the future volatility as well. We also expect each volatility component at different investment horizon to carry different information, which should again help to enhance the final forecasts. To be able to fully explore the decompositions, we use the extension of Realized Exponential GARCH model that can utilize multiple realized volatility measures introduced by \cite{hansen2012exponential}. The realized EGARCH model with $j=1,\ldots,J^m+1$ volatility components at different investment horizons estimated using JWTSRV in $x_{j,t}$ is
\begin{eqnarray}
	r_t &=& \sqrt{h_t} z_t,\\
	\log(h_t) &=& \omega + \beta\log(h_{t-1}) + \tau (z_{t-1}) + \mathbf{\gamma}' u_{t-1},\\
	\log(x_{j,t}) &=& \xi_j + \phi_j \log(h_t) + \delta_{(j)} (z_t) + u_{j,t},
\end{eqnarray}
where $z_t\sim N(0,1)$, and $u_t\sim N(0,\Sigma)$ are mutually and serially independent, and $u_t=(u_{1,t},\ldots,u_{j,t})'$, and $\tau (z_{t})=\tau_1 z_t+\tau_2 (z^2_t-1)$, and $\delta_{(j)}(z_t)=\delta_{j,1} z_t+\delta_{j,2} (z^2_t-1)$. 

Note that the model is different as the $\log(h_t)$ equation has the $u_{t-1}$ instead of realized measure, and includes leverage function. For the case when $j=1$, model is equivalent to the previous one, and by simple substitution, we can obtain the relation of parameters directly \citep{hansen2012exponential}. Hence the model with multiple equations is just a generalization of the previous work, which allows us to fully utilize the decomposed volatility into several investment horizons, and so parameters in vector $\gamma'$ will provide a good guide for significance of various investment horizons on volatility forecasts.

All the models are estimated by quasi-maximum likelihood farmework (QMLE) and can be easily generalized by assuming different distributions of $z_t$ and $u_t$. \cite{Hansen2011} provide the asymptotic properties of the QMLE, while \citep{hansen2012exponential} extend it to the framework with multiple realized measures, although the work is currently unfinished. The quasi log-likelihood function is given by
\begin{equation}
\ell(r,x;\theta, \Sigma) = -\frac{1}{2} \sum_{t=1}^T \left( \underbrace{\log (2 \pi) + \log(h_t) + z_t^2}_{=\ell(r)} + \underbrace{K \log (2 \pi) + \log (|\Sigma|) + u_t' \Sigma^{-1} u_t}_{=\ell(x|r)} \right),
\end{equation}
where $\theta$ holds set of parameters to be estimated by maximizing the quasi log-likelihood with respect to $\theta$ and $\Sigma$. The log-likelihood can be divided in two according to the contribution of realized measures to the log-likelihood value, $\ell(x|r)$ and contribution of returns, $\ell(r)$. In the empirical analysis, we report the two values as we use conventional GARCH model as a benchmark, so we are able to compare the fits. It is again straightforward to obtain multiple-period-ahead point forecasts using estimated parameters. For the details, see for example \cite{lunde2013modeling}

\subsection{Generalized Autoregressive Score Model with Realized Measures of Volatility and Jumps}

Recently introduced observation-driven estimation framework of Generalized Autoregressive Score (GAS) models due to \cite{creal2013generalized} has recently gained considerable popularity. \cite{huang2014generalized} propose a new observation-driven time-varying parameter Realized GARCH, in which the dynamic latent factor is updated by the scaled local density score as a function of past daily returns and realized variance. The new framework is robust to extreme outliers in observations, hence it may serve as a robustness check to our modeling strategy. We again add jumps to the original model, obtaining Realized Jump GAS Model as
\begin{eqnarray}
\label{RJ-GAS}
r_t &=&\sqrt{h_t} z_t, \\
\log(x_t)&=& \xi+\phi \log(h_t)+d_1(z_t^2-1)+d_2z_t+\sigma u_t,\\
\log(h_{t+1})&=& \omega+\alpha \mathcal{S}_t\Delta_t+\beta \log(h_t) +\gamma_J \log(1+JV_{t-1}),
\end{eqnarray}
where $x_t$ and $JV_t$ are estimated continuous and jump components of quadratic variation using BV, MedRV, or JWTSRV estimators, and $z_t$ and $u_t$ come from Gaussian normal distribution and are mutually independent. $d_1(z_t^2-1)+d_2z_t$ is leverage function that introduces dependence between the return shock and volatility shock. The main change in comparison to previous models is in the dynamics of the latent volatility, driven by the dynamic score, where $\Delta_t=\partial \ln p\left(r_t,\log(x_t)|\mathcal{F}_{t-1};\log(h_t),\theta\right) /\partial \log(h_t)$ is the conditional score at time $t$ and $\mathcal{S}_t=-E_{t-1}[\partial^2 \ln p\left(r_t,\log(x_t)|\mathcal{F}_{t-1};\log(h_t),\theta\right)/\partial^2 \log(h_t)^2]^{-1}$ is the scaling matrix. Analogously to the QMLE framework, likelihood can be separated to two parts, which we report in order to be able to compare the fits. Assuming both $z_t$, and $u_t$ follow independent standardized normal distributions, dynamic score reduces to $\Delta_t =  \frac{1}{2} ( z_t^2 - 1 ) + \frac{1}{\sigma} u_t \left( \phi + d_1 z_t^2 - \frac{d_2}{2} z_t \right)$, $\mathcal{S}_t^{-1}= \frac{1}{\sigma^2} \left( \phi^2 + 3 d_1^2 + \frac{d_2^2}{4} - 2 d_1 \phi \right) + \frac{1}{2}$.
Assuming exogenous ARMA structure for jumps, multiple-period-ahead forecasts are again obtained readily.

\subsection{Forecast evaluation using different realized variance measures}

To test significant differences of competing models, we use the Model Confidence Set (MCS) methodology of \cite{hansen2011model}. Given a set of forecasting models, $\mathcal{M}_{0}$, we identify the model confidence set $\widehat{\mathcal{M}}^*_{1-\alpha} \subset \mathcal{M}_{0}$, which is the set of models that contain the best forecasting model given a level of confidence $\alpha$. For a given model $i \in \mathcal{M}_{0}$, the $p$-value is the threshold confidence level. Model $i$ belongs to the MCS only if $\widehat{p}_i \ge \alpha$. MCS methodology repeatedly tests the null hypothesis of equal forecasting accuracy 
$$H_{0,\mathcal{M}}:E[L_{i,t}-L_{j,t}]=0, \hspace{1cm} \text{for all } i,j\in\mathcal{M}$$ 
with $L_{i,t}$ being an appropriate loss function of the $i$-th model. Starting with the full set of models, $\mathcal{M}=\mathcal{M}_0$, this procedure sequentially eliminates the worst-performing model from $\mathcal{M}$ when the null is rejected. The surviving set of models then belong to the model confidence set $\widehat{\mathcal{M}}^*_{1-\alpha}$. Following \cite{hansen2011model}, we implement the MCS using a stationary bootstrap with an average block length of 20 days.\footnote{We have used different block lengths, including the ones depending on the forecasting horizons, to assess the robustness of the results, without any change in the final results. These results are available from the authors upon request.} Two robust loss functions, mean square error (MSE) and QLIKEare used in the MCS \citep{Patton2011},while root mean square error (RMSE) is reported in the Tables.

\section{Empirical application: Does decomposition bring any improvement in volatility forecasting?}

\subsection{Data description}
Foreign exchange future contracts are traded on the Chicago Mercantile Exchange (CME) on a 24-hour basis. As these markets are among the most liquid, they are suitable for analysis of high-frequency data. We will estimate the realized volatility of British pound (GBP), Swiss franc (CHF) and euro (EUR) futures. All contracts are quoted in the unit value of the foreign currency in US dollars. It is advantageous to use currency futures data for the analysis instead of spot currency prices, as they embed interest rate differentials and do not suffer from additional microstructure noise coming from over-the-counter trading. The cleaned data are available from Tick Data, Inc.\footnote{http://www.tickdata.com/}

It is important to look first at the changes in the trading system before we proceed with the estimation on the data. In August 2003, for example, the CME launched the Globex trading platform, and for the first time ever in a single month, the trading volume on the electronic trading platform exceeded 1 million contracts every day. On Monday, December 18, 2006, the CME Globex(R) electronic trading platform started offering nearly continuous trading. More precisely, the trading cycle became 23 hours a day (from 5:00 pm on the previous day until 4:00 pm on current day, with a one-hour break in continuous trading), from 5:00 pm on Sunday until 4:00 pm on Friday. These changes certainly had a dramatic impact on trading activity and the amount of information available, resulting in difficulties in comparing the estimators on the pre-2003 data, the 2003--2006 data and the post--2006 data. For this reason, we restrict our analysis to a sample period extending from January 2, 2007 through August 20, 2014, which contains the most recent financial crisis. The futures contracts we use are automatically rolled over to provide continuous price records, so we do not have to deal with different maturities.

The tick-by-tick transactions are recorded in Chicago Time, referred to as Central Standard Time (CST). Therefore, in a given day, trading activity starts at 5:00 pm CST in Asia, continues in Europe followed by North America, and finally closes at 4:00 pm in Australia. To exclude potential jumps due to the one-hour gap in trading, we redefine the day in accordance with the electronic trading system. Moreover, we eliminate transactions executed on Saturdays and Sundays, US federal holidays, December 24 to 26, and December 31 to January 2, because of the low activity on these days, which could lead to estimation bias. Finally, we are left with 1902 days in the sample. Looking more deeply at higher frequencies, we find a large amount of multiple transactions happening exactly at the same time stamp. We use the arithmetic average for all observations with the same time stamp.

Having prepared the data, we can estimate the integrated volatility using different estimators and use them within proposed forecasting framework. For each futures contract, the daily quadratic variation is estimated using the realized variance estimator. Integrated variance and jump variation are estimated with the bipower variation, median estimator, and finally our jump wavelet two-scale realized variance estimator. All the estimators are adjusted for small sample bias. For convenience, we refer to the estimators in the description of the results as RV, BV, MedRV and JWTSRV, respectively, while the BV, MedRV, and JWTSRV estimators are used for decomposition of continuous and jump part of quadratic variation, and JWTSRV for decomposition to various investment horizons. We use sampling frequency of 5-minutes.

\begin{figure}[h]
   \centering
   \includegraphics[width=\textwidth]{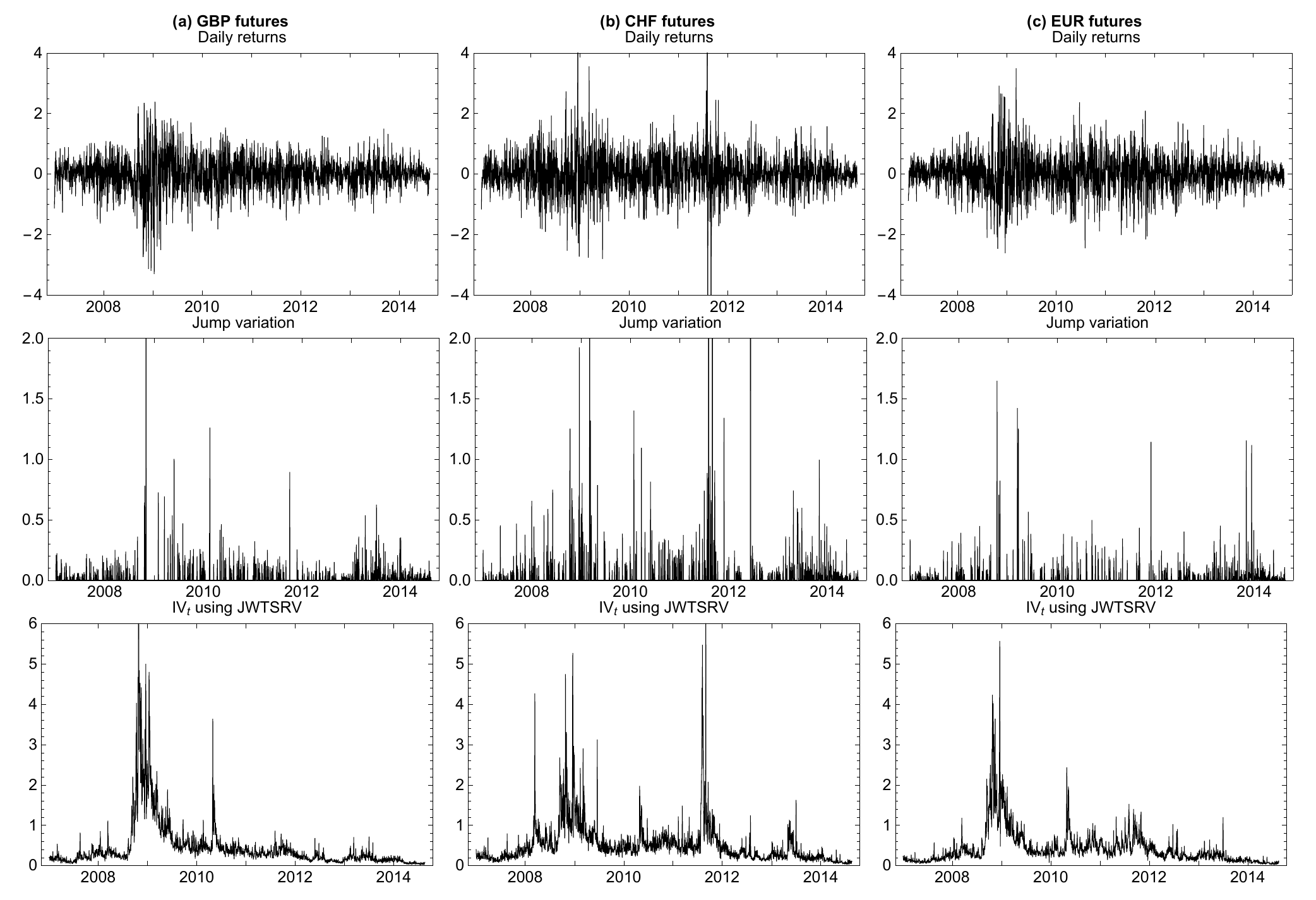} 
   \caption{Daily returns, estimated jump variation and $IV_t$ estimated by JWTSRV for (a) GBP, (b) CHF and (c) EUR futures.}
   \label{plotsBP}
\end{figure}

\begin{figure}
   \centering
   \includegraphics[width=\textwidth]{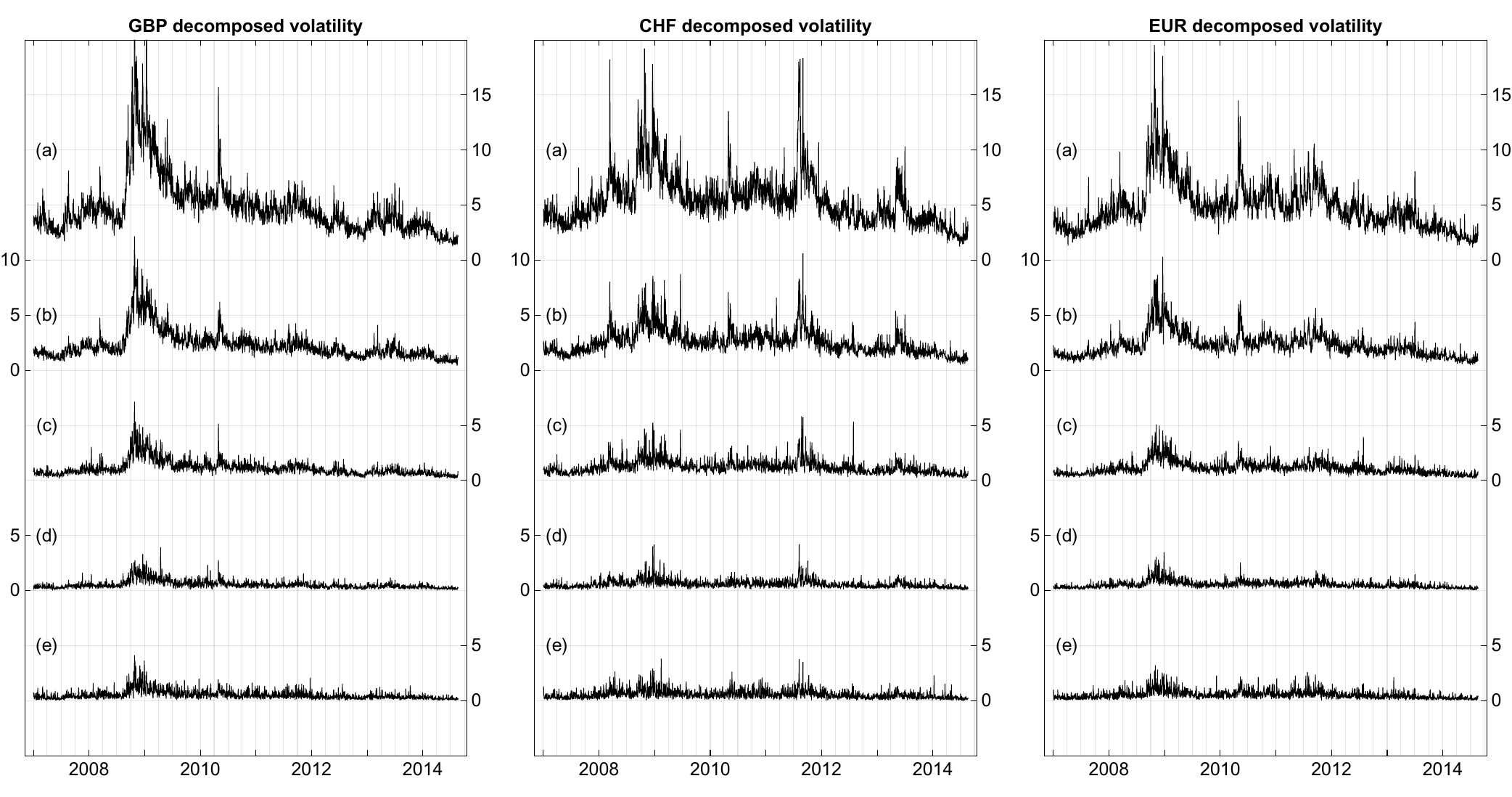} 
   \caption{Decomposed annualized volatility (by 252 days) of GBP, CHF and EUR futures using JWTSRV, (a) volatility on investment horizon of 10 minutes, (b) volatility on investment horizon of 20 minutes, (c) volatility on investment horizon of 40 minutes, (d) volatility on investment horizon of 80 minutes, (e) volatility on investment horizon up to 1 day. Note that sum of components (a), (b), (c), (d) and (e) give total volatility.} 
   \label{volatilityplot}
\end{figure}

The decomposition of volatility into the continuous and jump part is depicted by Figure \ref{plotsBP}, which provide the returns, estimated jump and finally integrated variance components using JWTSRV estimator for all three futures pairs. Figure \ref{volatilityplot} shows the further decomposition into several investment horizons. For better illustration, we annualize the square root of the integrated variance in order to get the annualized volatility and we compute the components of the volatility on several investment horizons. Figure \ref{volatilityplot} (a) to (e) show  the investment horizons of 10 minutes, 20 minutes, 40 minutes, 80 minutes and up to 1 day, respectively. It is very interesting that most of the volatility (around 50\%) comes from the 10-minute investment horizon which is a new empirical insight. Moreover, the longer the investment horizon, the lower the contribution of the variance to the total variation. 

\subsection{In-sample fits}

The main results of estimation and forecasting are presented in this section. The estimation strategy is as follows. For each of three forex futures considered, namely GBP, CHF and EUR, we first estimate benchmark GARCH(1,1) model. Then, we estimate the Realized GARCH (1,1) with RV, which will serve as a benchmark model to our Realized Jump GARCH(1,1) with BV, MedRV, and JWTSRV. All these models are estimated using QMLE and GAS model frameworks. Finally, we add Realized GARCH model with multiple JWTSRV components to see the impact of investment horizons on forecasts. 

Tables \ref{tab:RealizedGARCHBP}, \ref{tab:RealizedGARCHCHF} and \ref{tab:RealizedGARCHEUR} contain in-sample fits for GBP futures, CHF futures and EUR futures on the full sample respectively. By observing partial log-likelihood $\ell(r)$, we can see immediately that all the Realized GARCH models bring significant improvement to the conventional GARCH(1,1) without high frequency realized measures, reported by the first column (in testing significance of the difference, we restrict ourselves to use simple log-likelihood ratio test). 

When we focus on comparison of Realized GARCH models, we can observe further significant differences. Our Jump-GARCH brings small improvements to the $l(r)$ consistent with the literature, but large improvements in terms of $l(x|r)$ when compared to the benchmark Realized GARCH with RV. As to the comparison of QMLE and GAS specifications,  original QMLE model outperforms GAS in terms of likelihood slightly. These observations hold for all three futures used in the study.

Further comparison of the Realized Jump-GARCH models with three different realized measures reveals that JWTSRV and MedRV largely outperform BV, with JWTSRV bringing largest gains for CHF futures, and MedRV winning the race for the rest. While log-likelihoods $\ell(r,x)$ uncover rather large differences between the models, parameter estimates for the different realized measures are very similar to each other, and are consistent with the estimates found in the literature. 

The most important parameter $\gamma_J$ is significantly different from zero for BV and MedRV estimators, but not for JWTSRV estimator. We explain this by more strict statistics for testing the null hypothesis of no jumps in comparison to MedRV and BV, while we use bootstrap, which corrects the statistics for small sample distortions. As pointed out by \cite{dovonon2014bootstrapping}, the differences my be quite severe. Even with this result, we can conclude that jumps bring significant improvement in the modeling and Realized Jump-GARCH(1,1) outperforms benchmark Realized GARCH.

Finally, we focus on the Realized GARCH model with multiple measures, where we use volatility decompositions to several investment horizons due to our JWTSRV measure, and also include RV representing full quadratic variation. We find $\gamma_j$ coefficients statistically different from zero for all three futures. This means that volatility further decomposed to several investment horizons carry significant contribution to the future latent volatility. Coefficient is largest at the first scale, following the second, and the rest. This points us to the result that mainly volatility from highest frequency impacts the future volatility. 

Turning our attention to $\phi_j$, we can see that it is close to one (within standard errors) for all investment horizons. Note however how $\xi_j$ decreases with decreasing scale. This mirrors the different contributions of the energy (variance of each volatility at different investment horizon $j$ to total variance) to the latent volatility. From Eq. (\ref{jwtsrv}) we know, that volatility components at different horizons $j$ always sum up to the total volatility. But Realized GARCH model use logarithmic transforms, which do not hold this property. Hence, the expected value of the parameter $\xi_j$ will logically be a total constant minus $\log(1/2^j)$, as JWTSRV is simply sum of squared wavelet coefficients on intraday return, which is driven mainly by Brownian motion. This points us to the conclusion that the most of the information can be found in the high frequency part of the spetral density of returns.

\subsection{Multi-period-ahead forecasting results}

Motivated by a good in-sample performance of the models, we study if inclusion of jumps in the model improves the volatility forecasts in our newly proposed Realized Jump-GARCH models. We also wait to see if the model with multiple investment horizons improves volatility forecasts, and finally, it will be interesting to find out if the log-likelihood gains also translate to good forecasting performance of the models. 

We use all the Realized GARCH models to produce $h=\{1,5,10\}$-day-ahead forecasts based on rolling basis. Table \ref{tab:MCS} compares RMSE of all the models. To see if the forecasts are statistically different, we use the Model Confidence Set (MCS) with two robust loss functions, MSE and QLIKE. Models, which are included in the MCS with the use of both loss functions are highlighted in bold. In addition, we provide ranking of the models according to the both loss functions within MCS in the superscript. First number is ranking due to MSE, second one is ranking of the models due to QLIKE.

Turning to the results in the Table \ref{tab:MCS}, we can see that Realized GARCH model with multiple investment horizons is never rejected from the Model Confidence set by neither of the loss functions. Moreover, for GBP and EUR futures, it ranks as the best forecasting model with exception of forecasting horizon of 10 days, when it ranks as second according to QMLE. The model also delivers lowest RMSE of the forecasts, and ranks second to fifth with CHF futures outperformed mainly by GAS estimates.

Another model, which is never rejected by neither of the loss functions from MCS is the Realized Jump GAS model with our JWTSRV. For all three futures and all forecasting horizons, the model ranks as second best, to eighth best depending on the loss function. Similar results are delivered also with the use of MedRV, when the model often ranks third to fifth best, with one exception of forecasting EUR futures at horizons of five days. Realized Jump GAS model with BV is the third best model, as it is rejected from the MCS only for EUR futures with QLIKE loss function. 

Most of the Realized (Jump) GARCH models estimated using QMLE are rejected from MCS by one of the loss functions. The only exception is CHF forecasts at 10-day-ahead horizon, when the Realized Jump GARCH model with BV measure ranks best using both loss functions. 

Overall, the log-likelihood gains from QMLE estimates do not translate to better out-of-sample forecasts, as GAS outperforms the MLE models. Realized Jump-GARCH largely outperforms benchmark Realized GARCH with RV, and finally our multiple horizon model outperforms all the models delivering lowest loss functions most of the times. Thus jump variation as well as further decomposition of volatility to different scales bring significant improvement to the volatility forecasts in all tested forex futures. 

\section{Conclusion}

In this paper, we investigate how the decomposed integrated volatilities and jumps influence the future volatility using realized GARCH framework. Utilizing a jump wavelet two scale realized volatility estimator, which measures foreign exchange volatility in the time-frequency domain, we study the influence of intra-day investment horizons on daily volatility forecasts.

After the introduction of wavelet-based estimation of quadratic variation together with forecasting model, we compare our estimators to several most popular estimators, namely, realized variance, bipower variation, and median realized volatility in the forecasting exercise. Using several Realized GARCH specifications estimated by QMLE, GAS, and multiple realized measures, the wavelet-based estimator proves to bring significant improvement in the volatility forecasts. Models incorporating jumps improve forecasting ability significantly. Next, we find that while realized Jump GAS models do not outperform other models in terms of in-sample fits, they largely outperform the MLE-based estimates in the forecasts at all forecasting horizons. 

Concluding the empirical findings, we show that our wavelet-based estimators bring a significant improvement to the volatility estimation and forecasting. It also offers a new method of time-frequency modeling of realized volatility which helps us to better understand the dynamics of stock market behavior. Specifically, it uncovers that most of the volatility is created on higher frequencies.

{\footnotesize{
\setlength{\bibsep}{3pt}
\bibliographystyle{chicago}
\bibliography{thesis}
}}

\section*{Appendix:Tables}

\begin{landscape}
\begin{table}[tb]
	\tiny
	\caption{Results for the GBP futures: in-sample fits of GARCH(1,1), Realized GARCH(1,1) with RV, Realized Jump-GARCH with BV, MedRV, and JWTSRV estimated using MLE (Realized (Jump) GARCH) and GAS (Realized (Jump) GAS), and finally Realized GARCH with multiple $\widehat{IV}_{j,t}^{(JWTSRV)}$ volatility decompositions on different investment horizons. Robust standard errors are reported in parentheses.}

	\begin{center}
		\begin{tabular}{lccccccccclcccccc}
		\toprule
		 & GARCH & \multicolumn{4}{c}{Realized (Jump) GARCH} & \multicolumn{4}{c}{Realized (Jump) GAS} & & \multicolumn{6}{c}{Realized GARCH with multiple JWTSRV$_j$} \\
		\cmidrule(r){2-2} \cmidrule(r){3-6} \cmidrule(r){7-10} \cmidrule(r){12-17}
		  &  & RV & BV & MedRV & JWTSRV & RV & BV & MedRV & JWTSRV & & RV & $j=1$ & $j=2$ & $j=3$ & $j=4$ & $j=5$\\
		 \cmidrule(r){3-6} \cmidrule(r){7-10} \cmidrule(r){12-17}
 $\omega$ &  0.092  &  0.040  &  0.095  &  0.122  &  0.161  &  0.013  &  0.008  &  0.010  &  0.014  &  $\omega$ &  \multicolumn{6}{c}{0.019} \\ 
  &  (0.059)  & (0.030) & (0.034) & (0.036) & (0.036) & (0.008) & (0.009) & (0.009) & (0.010)  &      & \multicolumn{6}{c}{(0.009)} \\ 
  $\beta$ &  0.951  &  0.757  &  0.731  &  0.716  &  0.707  &  0.996  &  0.996  &  0.996  &  0.995   & $\beta$&  \multicolumn{6}{c}{0.994} \\ 
   &  (0.007)  & (0.018) & (0.020) & (0.021) & (0.021) & (0.002) & (0.003) & (0.003) & (0.003) &  & \multicolumn{6}{c}{(0.003)} \\ 
  $\gamma$&  0.046  &  0.224  &  0.235  &  0.244  &  0.242  &  0.237  &  0.261  &  0.275  &  0.283 &  $\gamma_j$ &  -0.011  &  0.128  &  0.050  &  0.008  &  0.023  &  0.013\\ 
   &  (0.007)  & (0.018) & (0.019) & (0.020) & (0.020) & (0.017) & (0.019) & (0.020) & (0.020)  & & (0.021)  &  (0.029)  &  (0.024)  &  (0.016)  &  (0.011)  &  (0.008)\\ 
  $\gamma_J$ &         &         &  0.015  &  0.008  &  0.002  &         &  0.017  &  0.009  &  0.002  &        \\ 
   &         &         & (0.005) & (0.004) & (0.004) &         & (0.005) & (0.005) & (0.005)  &        \\ 
   &  & & & & & & & &  &  $\tau_1$ & \multicolumn{6}{c}{-0.027} \\ 
  &  & & & & & & & &  &  & \multicolumn{6}{c}{(0.005)} \\ 
   &  & & & & & & & &  & $\tau_2$ & \multicolumn{6}{c}{0.032} \\ 
  &  & & & & & & & &  &  & \multicolumn{6}{c}{(0.004)} \\ 
    $\xi$ &  &  -0.128  &  -0.368  &  -0.455  &  -0.602  &  -0.009  &  -0.229  &  -0.320  &  -0.450  & $\xi_j$ & -0.273  &  -1.252  &  -2.052  &  -2.875  &  -3.659  &  -3.876 \\ 
   &  &  (0.131)  &  (0.145)  &  (0.153)  &  (0.155)  &  (0.064)  &  (0.136)  &  (0.143)  &  (0.145)  & & (0.125)  &  (0.134)  &  (0.136)  &  (0.140)  &  (0.144)  &  (0.139) \\ 
   $\phi$ &  &  1.070  &  1.127  &  1.147  &  1.189  &  1.040  &  1.091  &  1.113  &  1.150  &  $\phi_j$ &1.112  &  1.168  &  1.180  &  1.199  &  1.198  &  1.231 \\ 
   &  &  (0.039)  &  (0.043)  &  (0.045)  &  (0.046)  &  (0.021)  &  (0.041)  &  (0.043)  &  (0.043)  & & (0.036)  &  (0.039)  &  (0.040)  &  (0.041)  &  (0.042)  &  (0.041) \\ 
  $\tau_1/d_1$  &  &  -0.017  &  -0.024  &  -0.025  &  -0.028  &  0.079  &  0.064  &  0.056  &  0.058  & $\delta_{j,1}$& -0.014  &  -0.028  &  -0.026  &  -0.021  &  -0.020  &  -0.032 \\ 
   &  &  (0.008)  &  (0.007)  &  (0.007)  &  (0.007)  &  (0.005)  &  (0.005)  &  (0.005)  &  (0.005)  & & (0.008)  &  (0.007)  &  (0.008)  &  (0.010)  &  (0.013)  &  (0.015) \\ 
  $\tau_2/d_2$  &  &  0.087  &  0.072  &  0.066  &  0.068  &  -0.003  &  -0.008  &  -0.009  &  -0.010  &  $\delta_{j,2}$&0.087  &  0.060  &  0.065  &  0.074  &  0.087  &  0.209 \\ 
   &  &  (0.006)  &  (0.005)  &  (0.005)  &  (0.005)  &  (0.007)  &  (0.007)  &  (0.007)  &  (0.007)  &  &(0.006)  &  (0.005)  &  (0.006)  &  (0.007)  &  (0.009)  &  (0.012) \\ 
		\cmidrule(r){2-2} \cmidrule(r){3-6} \cmidrule(r){7-10} \cmidrule(r){12-17}
  $\ell(x|r)$  &     &  -613  &  -533  &  -486  &  -506  &  -625  &  -546  &  -501  &  -523  &  &\multicolumn{6}{c}{-2543} \\ 
  $\ell(r)$ &  -5849  &  -5825  &  -5824  &  -5823  &  -5825  &  -5825  &  -5825  &  -5823  &  -5825  & & \multicolumn{6}{c}{-5825} \\ 
 		\bottomrule
		\end{tabular}
	\end{center}
		\label{tab:RealizedGARCHBP}
\end{table}
\end{landscape}

\begin{landscape}
\begin{table}[tb]
	\tiny
	\caption{Results for the CHF futures: in-sample fits of GARCH(1,1), Realized GARCH(1,1) with RV, Realized Jump-GARCH with BV, MedRV, and JWTSRV estimated using MLE (Realized (Jump) GARCH) and GAS (Realized (Jump) GAS), and finally Realized GARCH with multiple $\widehat{IV}_{j,t}^{(JWTSRV)}$ volatility decompositions on different investment horizons. Robust standard errors are reported in parentheses.}

	\begin{center}
		\begin{tabular}{lccccccccclcccccc}
		\toprule
		 & GARCH & \multicolumn{4}{c}{Realized (Jump) GARCH} & \multicolumn{4}{c}{Realized (Jump) GAS} & & \multicolumn{6}{c}{Realized GARCH with multiple JWTSRV$_j$} \\
		\cmidrule(r){2-2} \cmidrule(r){3-6} \cmidrule(r){7-10} \cmidrule(r){12-17}
		  &  & RV & BV & MedRV & JWTSRV & RV & BV & MedRV & JWTSRV & & RV & $j=1$ & $j=2$ & $j=3$ & $j=4$ & $j=5$\\
		 \cmidrule(r){3-6} \cmidrule(r){7-10} \cmidrule(r){12-17}
 $\omega$ & 0.092  &  -0.140  &  -0.095  &  -0.108  &  -0.098  &  0.018  &  0.016  &  0.019  &  0.023  & $\omega$ & \multicolumn{6}{c}{0.033} \\ 
   &  (0.087)  & (0.031) & (0.034) & (0.038) & (0.036) & (0.012) & (0.013) & (0.013) & (0.013) &  & \multicolumn{6}{c}{(0.014)} \\ 
  $\beta$ &  0.936  &  0.760  &  0.729  &  0.707  &  0.719  &  0.995  &  0.994  &  0.993  &  0.994  &  $\beta$& \multicolumn{6}{c}{0.991} \\ 
   &  (0.008)  & (0.018) & (0.019) & (0.020) & (0.020) & (0.003) & (0.003) & (0.003) & (0.003) & & \multicolumn{6}{c}{(0.004)} \\ 
 $\gamma$ &  0.066  &  0.276  &  0.297  &  0.324  &  0.311  &  0.235  &  0.267  &  0.275  &  0.268  & $\gamma_j$ & 0.030  &  0.262  &  -0.020  &  0.068  &  -0.010  &  0.031 \\ 
   &  (0.008)  & (0.021) & (0.022) & (0.023) & (0.022) & (0.018) & (0.019) & (0.019) & (0.019) &  & (0.033)  &  (0.043)  &  (0.037)  &  (0.026)  &  (0.018)  &  (0.012) \\ 
  $\gamma_J$ &         &         &  0.016  &  0.012  &  0.004  &         &  0.017  &  0.013  &  0.002  &        \\ 
   &         &         &  (0.007) & (0.006) & (0.005) &         & (0.007) & (0.006) & (0.006)  &        \\ 
   &  & & & & & & & &  & $\tau_1$ & \multicolumn{6}{c}{0.021} \\ 
  &  & & & & & & & &  &  & \multicolumn{6}{c}{(0.007)} \\ 
   &  & & & & & & & &  & $\tau_2$ & \multicolumn{6}{c}{0.030} \\ 
  &  & & & & & & & &  &  & \multicolumn{6}{c}{(0.005)} \\ 
  $\xi$ &  &  0.602  &  0.389  &  0.412  &  0.399  &  0.786  &  0.572  &  0.571  &  0.536  &  $\xi_j$ & 0.677  &  -0.200  &  -0.964  &  -1.711  &  -2.534  &  -2.802 \\ 
   &  &  (0.092)  &  (0.099)  &  (0.102)  &  (0.101)  &  (0.075)  &  (0.084)  &  (0.094)  &  (0.093)  & & (0.079)  &  (0.083)  &  (0.085)  &  (0.088)  &  (0.096)  &  (0.101) \\ 
  $\phi$ &  &  0.844  &  0.885  &  0.875  &  0.878  &  0.791  &  0.833  &  0.828  &  0.837  $\phi_j$& &  0.823  &  0.846  &  0.846  &  0.842  &  0.850  &  0.886 \\ 
   &  &  (0.024)  &  (0.026)  &  (0.027)  &  (0.027)  &  (0.020)  &  (0.022)  &  (0.025)  &  (0.025)  &  &(0.021)  &  (0.022)  &  (0.022)  &  (0.023)  &  (0.026)  &  (0.027) \\ 
  $\tau_1/d_1$ &  &  0.030  &  0.020  &  0.024  &  0.019  &  0.083  &  0.077  &  0.059  &  0.059  & $\delta_{j,1}$ & 0.028  &  0.022  &  0.016  &  0.011  &  0.025  &  0.001 \\ 
   &  &  (0.009)  &  (0.008)  &  (0.008)  &  (0.008)  &  (0.005)  &  (0.005)  &  (0.004)  &  (0.004)  &  &(0.008)  &  (0.007)  &  (0.008)  &  (0.010)  &  (0.012)  &  (0.007) \\ 
  $\tau_2/d_2$ &  &  0.098  &  0.089  &  0.073  &  0.074  &  0.030  &  0.020  &  0.025  &  0.017  &  $\delta_{j,2}$ & 0.100  &  0.071  &  0.075  &  0.084  &  0.087  &  0.185 \\ 
   &  &  (0.006)  &  (0.006)  &  (0.005)  &  (0.005)  &  (0.008)  &  (0.008)  &  (0.008)  &  (0.008)  & & (0.006)  &  (0.005)  &  (0.006)  &  (0.007)  &  (0.008)  &  (0.012) \\ 
		\cmidrule(r){2-2} \cmidrule(r){3-6} \cmidrule(r){7-10} \cmidrule(r){12-17}
   $\ell(x|r)$  &     &  -825  &  -753  &  -718  &  -688  &  -832  &  -760  &  -731  &  -705  &&  \multicolumn{6}{c}{-3286} \\ 
  $\ell(r)$  &  -6199  &  -6164  &  -6164  &  -6163  &  -6164  & - 6173  &  -6170  &  -6170  &  -6169  &&  \multicolumn{6}{c}{-6167} \\ 
 		\bottomrule
		\end{tabular}
	\end{center}
		\label{tab:RealizedGARCHCHF}

\end{table}
\end{landscape}

\begin{landscape}
\begin{table}[tb]
	\tiny
	\caption{Results for the EUR futures: in-sample fits of GARCH(1,1), Realized GARCH(1,1) with RV, Realized Jump-GARCH with BV, MedRV, and JWTSRV estimated using MLE (Realized (Jump) GARCH) and GAS (Realized (Jump) GAS), and finally Realized GARCH with multiple $\widehat{IV}_{j,t}^{(JWTSRV)}$ volatility decompositions on different investment horizons. Robust standard errors are reported in parentheses.}

	\begin{center}
		\begin{tabular}{lccccccccclcccccc}
		\toprule
		 & GARCH & \multicolumn{4}{c}{Realized (Jump) GARCH} & \multicolumn{4}{c}{Realized (Jump) GAS} & & \multicolumn{6}{c}{Realized GARCH with multiple JWTSRV$_j$} \\
		\cmidrule(r){2-2} \cmidrule(r){3-6} \cmidrule(r){7-10} \cmidrule(r){12-17}
		  &  & RV & BV & MedRV & JWTSRV & RV & BV & MedRV & JWTSRV & & RV & $j=1$ & $j=2$ & $j=3$ & $j=4$ & $j=5$\\
		 \cmidrule(r){3-6} \cmidrule(r){7-10} \cmidrule(r){12-17}
 $\omega$ &  0.039  &  -0.063  &  -0.022  &  0.011  &  0.047  &  0.026  &  0.018  &  0.020  &  0.028  & $\omega$ &  \multicolumn{6}{c}{0.039} \\ 
  &  (0.048)  & (0.038) & (0.043) & (0.079) & (0.041) & (0.012) & (0.013) & (0.013) & (0.015) &  & \multicolumn{6}{c}{(0.013)} \\ 
  $\beta$ &  0.955  &  0.734  &  0.698  &  0.683  &  0.684  &  0.992  &  0.993  &  0.992  &  0.992  & $\beta$& \multicolumn{6}{c}{0.989} \\ 
   &  (0.006)  & (0.020) & (0.022) & (0.022) & (0.022) & (0.003) & (0.004) & (0.003) & (0.004) &  & \multicolumn{6}{c}{(0.004)} \\ 
  $\gamma$ &  0.045  &  0.286  &  0.313  &  0.321  &  0.311  &  0.260  &  0.296  &  0.308  &  0.308  &  $\gamma_j$ &-0.016  &  0.182  &  0.036  &  0.049  &  -0.008  &  0.041 \\ 
   &  (0.006)  & (0.024) & (0.025) & (0.029) & (0.024) & (0.020) & (0.021) & (0.021) & (0.021) &  & (0.028)  &  (0.037)  &  (0.032)  &  (0.022)  &  (0.016)  &  (0.011) \\ 
  $\gamma_J$ &         &         &  0.018  &  0.012  &  0.001  &         &  0.019  &  0.014  &  0.001 &        \\ 
   &         &         &  (0.008) & (0.007) & (0.003) &         & (0.008) & (0.007) & (0.009)  &        \\ 
   &  & & & & & & & &  & $\tau_1$ & \multicolumn{6}{c}{-0.027} \\ 
  &  & & & & & & & &  &  & \multicolumn{6}{c}{(0.006)} \\ 
   &  & & & & & & & &  & $\tau_2$ & \multicolumn{6}{c}{0.051} \\ 
  &  & & & & & & & &  &  & \multicolumn{6}{c}{(0.005)} \\ 
  $\xi$ &  &  0.302  &  0.122  &  0.030  &  -0.063  &  0.452  &  0.297  &  0.230  &  0.101  &  $\xi_j$ & 0.155  &  -0.866  &  -1.683  &  -2.434  &  -3.307  &  -3.456 \\ 
   &  &  (0.114)  &  (0.124)  &  (0.224)  &  (0.127)  &  (0.099)  &  (0.107)  &  (0.112)  &  (0.118)  & & (0.118)  &  (0.128)  &  (0.131)  &  (0.135)  &  (0.143)  &  (0.135) \\ 
  $\phi$ &  &  0.907  &  0.945  &  0.965  &  0.990  &  0.869  &  0.901  &  0.916  &  0.950  &  $\phi_j$& 0.948  &  1.015  &  1.032  &  1.034  &  1.064  &  1.077 \\ 
   &  &  (0.033)  &  (0.035)  &  (0.062)  &  (0.036)  &  (0.029)  &  (0.031)  &  (0.032)  &  (0.034)  & & (0.033)  &  (0.036)  &  (0.037)  &  (0.038)  &  (0.040)  &  (0.038) \\ 
  $\tau_1/d_1$  &  &  0.001  &  0.001  &  -0.004  &  -0.009  &  0.088  &  0.076  &  0.069  &  0.067  &  $\delta_{j,1}$ & 0.001  &  -0.007  &  -0.007  &  -0.002  &  0.014  &  -0.012 \\ 
   &  &  (0.005)  &  (0.005)  &  (0.008)  &  (0.008)  &  (0.006)  &  (0.006)  &  (0.005)  &  (0.005)  &  &(0.008)  &  (0.008)  &  (0.009)  &  (0.011)  &  (0.014)  &  (0.016) \\ 
  $\tau_2/d_2$ &  &  0.093  &  0.083  &  0.077  &  0.075  &  0.010  &  0.011  &  0.008  &  0.003  & $\delta_{j,2}$ & 0.093  &  0.072  &  0.076  &  0.073  &  0.075  &  0.207 \\ 
   &  &  (0.006)  &  (0.006)  &  (0.005)  &  (0.005)  &  (0.008)  &  (0.007)  &  (0.007)  &  (0.007)  & & (0.006)  &  (0.005)  &  (0.006)  &  (0.007)  &  (0.009)  &  (0.012) \\ 
		\cmidrule(r){2-2} \cmidrule(r){3-6} \cmidrule(r){7-10} \cmidrule(r){12-17}
   $\ell(x|r)$ &     &  -715  &  -655  &  -582  &  -600  &  -710  &  -651  &  -582  &  -601  & & \multicolumn{6}{c}{-2555} \\ 
  $\ell(r)$ &  -6016  &  -5994  &  -5993  &  -5992  &  -5990  &  -5996  &  -5996  &  -5994  &  -5992  & & \multicolumn{6}{c}{-5995} \\ 
 		\bottomrule
		\end{tabular}
	\end{center}
		\label{tab:RealizedGARCHEUR}
\end{table}
\end{landscape}

\begin{table}[h]
	\tiny
	\caption{RMSE $(\times 10^{-4})$ from all forecasts for the GBP, CHF, and EUR at different forecasting horizons $h=\{1,5,10\}$. Forecasts which fall into the 10\% Model Confidence Set (MCS) using both robust MSE and QLIKE loss functions are in bold. In addition, ranking of the models included in the MCS is provided in the superscript, first is ranking using MSE, second using QLIKE.}
	\begin{center}
		\begin{tabular}{lllllllllll}
		\toprule
		& \multicolumn{4}{c}{Realized (Jump) GARCH} & \multicolumn{4}{c}{Realized (Jump) GAS} & Multiple \\
		 \cmidrule(r){2-5} \cmidrule(r){6-9}  
		   & RV & BV & MedRV & JWTSRV & RV & BV & MedRV & JWTSRV & JWTSRV$_j$ \\
		 \cmidrule(r){2-5} \cmidrule(r){6-9}  \cmidrule(r){10-10}
		   \multicolumn{1}{l}{\textbf{GBP}} \\
		 $h=1$ & ${1.007}^{(9)}$ & ${1.003}^{(7)}$ & ${1.003}^{(6)}$ & ${1.004}^{(8)}$ & $\mathbf{0.998}^{(5,5)}$ & $\mathbf{0.988}^{(4,4)}$ & $\mathbf{0.986}^{(3,3)}$ & $\mathbf{0.982}^{(2,2)}$ & $\mathbf{0.972}^{(1,1)}$\\
  $h=5$ & $\mathbf{0.644}^{(6,5)}$ & $0.658$ & ${0.660}$ & ${0.661}$ & $\mathbf{0.643}^{(5,3)}$ & $\mathbf{0.639}^{(2,2)}$ & $\mathbf{0.641}^{(3,4)}$ & $\mathbf{0.642}^{(4,6)}$ & $\mathbf{0.633}^{(1,1)}$\\
  $h=10$ & $\mathbf{0.561}^{(2,1)}$ & $\mathbf{0.566}^{(3,3)}$ & $\mathbf{0.569}^{(5,4)}$ & $\mathbf{0.574}^{(8,7)}$ & $\mathbf{0.572}^{(7,6)}$ & $\mathbf{0.568}^{(4,5)}$ & $\mathbf{0.571}^{(6,8)}$ & $\mathbf{0.575}^{(9,9)}$ & $\mathbf{0.558}^{(1,2)}$\\
   \cmidrule(r){2-5} \cmidrule(r){6-9} \cmidrule(r){10-10}  
   		   \multicolumn{1}{l}{\textbf{CHF}} \\
$h=1$ & ${1.497}^{(3)}$ & ${1.459}^{(1)}$ & ${1.517}^{(6)}$ & ${1.496}^{(2)}$ & ${1.591}^{(3)}$ & $\mathbf{1.497}^{(4,1)}$ & $\mathbf{1.530}^{(7,5)}$ & $\mathbf{1.538}^{(8,2)}$ & $\mathbf{1.509}^{(5,4)}$\\
  $h=5$ & $\mathbf{1.087}^{(5,5)}$ & $\mathbf{1.027}^{(1,4)}$ & ${1.095}^{(6)}$ & ${1.073}^{(2)}$ & ${1.211}^{(6)}$ & $\mathbf{1.077}^{(3,1)}$ & $\mathbf{1.120}^{(7,7)}$ & $\mathbf{1.140}^{(8,3)}$ & $\mathbf{1.087}^{(4,2)}$\\
  $h=10$ & $\mathbf{1.155}^{(7,2)}$ & $\mathbf{1.079}^{(1,1)}$ & $\mathbf{1.119}^{(3,7)}$ & $\mathbf{1.125}^{(4,5)}$ & ${1.224}^{(9)}$ & $\mathbf{1.107}^{(2,4)}$ & $\mathbf{1.136}^{(6,8)}$ & $\mathbf{1.163}^{(8,6)}$ & $\mathbf{1.126}^{(5,3)}$\\
     \cmidrule(r){2-5} \cmidrule(r){6-9} \cmidrule(r){10-10}  
   		   \multicolumn{1}{l}{\textbf{EUR}} \\
 $h=1$ & ${1.280}^{(9)}$ & ${1.273}^{(6)}$ & ${1.253}^{(4)}$ & $\mathbf{1.230}^{(2,3)}$ & $\mathbf{1.279}^{(8,5)}$ & ${1.278}^{(7)}$ & $\mathbf{1.264}^{(5,4)}$ & $\mathbf{1.244}^{(3,2)}$ & $\mathbf{1.221}^{(1,1)}$\\ 
  $h=5$ &${0.981}$ & ${0.970}^{(8)}$ & ${0.943}^{(4)}$ & ${0.927}^{(2)}$ & ${0.960}^{(6)}$ & ${0.961}^{(7)}$ & ${0.946}^{(5)}$ & $\mathbf{0.930}^{(3,2)}$ & $\mathbf{0.911}^{(1,1)}$\\ 
  $h=10$ & ${0.974}$ & ${0.936}$ & ${0.894}^{(6)}$ & ${0.905}^{(3)}$ & ${0.876}^{(5)}$ & $0.870^{(7)}$ & $\mathbf{0.848}^{(3,4)}$ & $\mathbf{0.848}^{(2,1)}$ & $\mathbf{0.848}^{(1,2)}$\\
\bottomrule
		\end{tabular}
	\end{center}
		\label{tab:MCS}
\end{table}

\section*{Appendix: Figures}

\clearpage
\newpage
\appendix
\section{Wavelet transform}  \label{wt}

In this Appendix we briefly introduce basic ideas of wavelet transform. Let us begin with the  continuous wavelet transform which is a cornerstone of the wavelet analysis. Further we introduce a special form of discrete wavelet transform called the ``maximal overlap discrete wavelet transform'' (MODWT) that we use in empirical part. Following \cite{Daubechies1992} and \cite{Chui1992} we define doubly-indexed wavelet function -- a wavelet\footnote{An important conditions a wavelet function must fulfill is the admissibility condition: $C_{\psi }{\rm =}\int^\infty_0{}\frac{{\rm |}\Psi {\rm (}f{\rm )}{{\rm |}}^{{\rm 2}}}{f}df{\rm <}\infty$, where $\Psi {\rm (}f{\rm )}$ is the Fourier transform of a wavelet $\psi {\rm (.)}$. For more details about wavelet filer conditions see \cite{Daubechies1992}} as:
\begin{equation}
\psi_{j,k}\left(t\right)=\frac{1}{\sqrt{j}}\psi \left(\frac{t-k}{j}\right) \in L^{{2}}({\mathbb R}),
\end{equation}
where index $k$ determines the exact position of the wavelet in time, whereas  the scaling index $j$ controls how the wavelet is stretched or dilated, i.e., frequency resolution of the wavelet. The continuous wavelet transform, $W_{j,k}$, is a projection of a wavelet function $\psi_{j,k}$ onto the time series $y{\rm (}t{\rm )} \in L^{{\rm 2}}{\rm (}{\mathbb R}{\rm )}$:
\begin{equation}
\label{cwt} 
W_{j,k}=\int^\infty_\infty y(t)\overline{\psi}_{j,k}(t)dt.
\end{equation}
Hence, Eq.(\ref{cwt}) tranforms $y(t)$, time-domain process, into $W_{j,k}$ which is time-frequency (or time-scale) space, where $k$ is the position in time and $j$ corresponds to a specific frequency. 
Using the wavelet coefficients $W_{j,k}$ we can subsequently recover the time series $y(t)$ as follows:
\begin{equation} 
\label{GrindEQ__8_} 
y(t)=\frac{{\rm 1}}{C_{\psi }}\int^\infty_0\left[\int^\infty_{{\rm -}\infty}W_{j,k}{\psi }_{j,k}{\rm (}t{\rm )}dk\right]\frac{dj}{j^{{\rm 2}}}{\rm ,\ \ \ \ }k{\rm >}0. 
\end{equation}
The continuous wavelet transform preserves variance of the analyzed time series. It is an important property that allows us to work with the decomposed wavelet variance. Thus we can write:
\begin{equation}
\label{GrindEQ__9_} 
x^2=\frac{{\rm 1}}{C_{\psi }}\int^\infty_0{}\left[\int^\infty_{{\rm -}\infty}{}{\left|W_{j,k}\right|}^{{\rm 2}}dk\right]\frac{dj}{j^{{\rm 2}}}. 
\end{equation} 
For a more detailed introduction to continuous wavelet transform and wavelets, see \cite{Daubechies1992}, \cite{Chui1992}, and \cite{PercivalWalden2000}.

In empirical applications we work with discrete time series, thus some form of discretization is needed. The discrete wavelet transform (DWT)\footnote{For a definition and detailed discussion of the discrete wavelet transform, see \cite{Mallat98}, \cite{PercivalWalden2000}, and \cite{Gencay2002}.}, which is a parsimonious form of the continuous wavelet transform allows for an analysis of discrete time series where only a bounded number of scales is required. The discrete version of wavelet transform has, however, some serious limitation that make its application to real time series rather difficult. These are mainly the sample size restriction to the power of two and the starting point sensitivity of the wavelet transform. 

\subsection{Maximal overlap discrete wavelet transform} \label{modwt}

The MODWT is in some cases superior to the DWT for empirical data analysis. For example, the problem of sample length restriction is connected with downsampling procedure of the DWT. However, the construction of MODWT does not use downsampling, thus vectors of the wavelet coefficients at all scales have equal length, corresponding to the length of transformed time series. As a consequence, the MODWT is not restricted to any sample size. In addition, the MODWT is a translation-invariant; therefore, it is not sensitive to the choice of the starting point of the examined time series. Similarly as the CWT, the MODWT wavelet and scaling coefficients can be used for analysis of variance of a time series in the time-frequency domain. Statistical properties of the MODWT variance estimators for non-stationary and non-Gaussian processes are discussed in detail in \cite{Serroukh2000}. For additional details on the MODWT, see \cite{Mallat98} and \cite{PercivalWalden2000}.

For computation of the MODWT coefficients we apply the pyramid algorithm of \cite{Mallat98}. The procedure is based on filtering time series with MODWT wavelet filters; the filtered time series is then filtered again in a subsequent stages to obtain other wavelet scales. These scales contain information localized at corresponding frequency bands of analyzed time series. 

Let us briefly introduce the pyramid algorithm. In the first stage, the wavelet coefficients are obtained via circular filtering of time series $y_t$ using the MODWT wavelet and scaling filters $h_{1,l}$ and $g_{1,l}$ \citep{PercivalWalden2000}:
\begin{equation} 
\label{GrindEQ__13_} 
\mathcal{W}_{1,k}\equiv \sum^{L{\rm -}{\rm 1}}_{l{\rm =0}}{}h_{1,l}\ y_{k-l \ modN,\ \ \ \ }\mathcal{V}_{1,k}\equiv \sum^{L{\rm -}{\rm 1}}_{l{\rm =0}}{}g_{1,l}\ y_{k-l \ modN}, 
\end{equation}
where $L_j=2^{j-1}\left(L-1\right)+1$ defines a width of the wavelet and scaling filters.\footnote{For more information about wavelet filters see for example \cite{PercivalWalden2000}.} After the first stage we obtain the wavelet and scaling coefficients at the first scale ($j=1$). The algorithm continues with the second stage where instead of $y_t$ we filter the sequence of scaling coefficients from the first stage $\mathcal{V}_{1,k}$, using the MODWT wavelet and scaling filters $h_{2,l}$ and $g_{2,l}$ for the second scale, i.e.,
\begin{equation} 
\label{GrindEQ__14_} 
\mathcal{W}_{2,k}\equiv \sum^{L{\rm -}{\rm 1}}_{l{\rm =0}}h_{2,l}\mathcal{V}_{1,k-l \ modN,\ \ \ \ }\mathcal{V}_{2,k}\equiv \sum^{L{\rm -}{\rm 1}}_{l{\rm =0}}{}g_{2,l}\mathcal{V}_{1,k-l \ modN.} 
\end{equation}
We may continue with more stages until the level of decomposition is $j\le log_{2}{\rm (}N{\rm )}$. For example, in case we need two levels of decomposition, i.e, we apply two stages, we obtain two vectors of wavelet coefficients; $\mathcal{W}_{1,k}$, $\mathcal{W}_{2,k}$ and a vector of the scaling coefficients at scale two $\mathcal{V}_{2,k}$, where $k=0,1,\dots ,N-1$. Vectors of wavelet and scaling coefficients reflect variations at specific frequency bands. Generally, $\mathcal{W}_{j,.}$ represents a frequency band $f [1/2^{j +1}{\rm ,1/}{{\rm 2}}^j{\rm ]}$, whereas $\mathcal{V}_{j,.}$ represents a frequency band $f [0,1/2^{j+1}]$.

\end{document}